\documentclass[twocolumn,showpacs,preprintnumbers,amsmath,amssymb]{revtex4}

\usepackage{graphicx}
\usepackage{dcolumn}
\usepackage{bm}
\usepackage {longtable}

\begin{document}

\preprint{}

\title{Investigation of cold Rb Rydberg atoms in a magneto-optical trap }
\author{D.~B.~Tretyakov$^1$}
\author{I.~I.~Beterov$^1$}
\author{V.~M.~Entin$^1$}
\author{I.~I.~Ryabtsev$^1$}
  \email{ryabtsev@isp.nsc.ru}
\author{P.~L.~Chapovsky$^2$}

\affiliation{$^1$Institute of Semiconductor Physics, Prospekt Lavrentyeva 13, 630090 Novosibirsk, Russia\\ $^2$Institute of Automation and Electrometry, Prospekt Koptyuga 1, 630090 Novosibirsk, Russia }

\date{September 4, 2008}

\begin{abstract}
We present our results on the experiments with cold Rb Rydberg atoms in a 
magneto-optical trap (MOT). Characteristic features of our experiment were 
the excitation of Rydberg atoms in a small volume within the cold atom cloud 
and sorting of the measured signals and spectra over the number of 
registered Rydberg atoms. We have measured the effective lifetime of the 
Rydberg state 37\textit{P}, as well as its polarizability in a weak electric 
field. The results are in good agreement with the theoretical calculations. 
We have shown that localization of the small excitation volume around the 
zero-magnetic-field point makes possible to increase the spectral resolution 
and to obtain narrow microwave resonances in Rydberg atoms without switching 
off the MOT quadrupole magnetic field. We have measured the dependence of 
the amplitude of the dipole-dipole interaction resonances on the number of 
Rydberg atoms, which has a linear character and agrees with the theory 
for weak dipole-dipole interaction.
\end{abstract}

 \maketitle

\section{INTRODUCTION}

In recent years the cold atoms in highly excited Rydberg states have 
attracted increasing attention due to the fact that the gas consisting of 
almost frozen Rydberg atoms behaves in a way similar to amorphous solids, in 
which the strong collective inter-atomic interactions can lead to broadening 
and shift of the spectral lines [1-3], as well as to ionization of atoms 
and formation of ultracold plasma [4].

Long-range interactions of cold Rydberg atoms are of particular interest due 
to their possible application to quantum computing. The basic ideas were 
formulated in paper [5] for two-qubit logical gates and in [6] for the 
effect of dipole blockade at the collective excitation of an ensemble of 
cold atoms. Two-qubit operations can be realized at short excitation of two 
adjacent Rydberg atoms, whose interaction leads to variation in the phase of 
the collective wave function. The dipole blockade is an effect of Rydberg 
excitation of only one atom in the ensemble because the resonances of 
multiatom excitation are shifted due to the interatomic interactions. Both 
effects can be realized using the dipole-dipole interaction of Rydberg 
atoms. However, until now they have not been observed experimentally for a 
small number of atoms, and thus become a key issue in the possible 
realization of a quantum computer based on neutral atoms.

Thanks to their large dipole moments, Rydberg atoms can effectively interact 
at distances of several and even tens of microns. The two pioneering 
experiments [7,8] report on the influence of van der Waals interaction on 
the probability and spectra of the optical transitions with principal 
quantum number $n > 60$. The effect is interpreted as "van der Waals 
blockade" and is analogous to the dipole blockade. In papers [9-12] a 
resonance microwave radiation was used to control the dipole-dipole 
interaction.

The dynamics of the signals of the resonant dipole-dipole interaction was 
investigated in paper [13], where the dependences of the Rydberg states 
population on the interaction time and number of atoms have been studied. 
Finally, the experimental works [14,15] describe the first attempts to 
observe the effect of dipole blockade. The resonant interaction caused a 
decrease in the Rydberg atom excitation probability of several times. 
However, dipole blockade effect with excitation of a single atom has not yet 
been reported.

Despite the large number of publications devoted to investigation of 
long-range interactions of Rydberg atoms, many aspects of these interactions 
remain unstudied, especially for a small number of atoms. This is because of 
the fact that Rydberg atoms were mainly registered using micro-channel-plate 
detectors, which do not provide resolution over the number of atoms. At the 
same time, in a series of experiments performed by our group [16,17], for 
registration of Na Rydberg atoms in a thermal beam we used a channel 
electron multiplier (channeltron) VEU-6, which allowed registration and 
sorting of signals according to the number of detected atoms (from 1 to 5). 
In the present work we apply for the first time this approach to cold 
Rydberg atoms.

Further we present the first results of our experiments with cold Rb Rydberg 
atoms. A feature of these experiments is the excitation of a few Rydberg 
atoms in a small volume within the cold atomic cloud in a MOT. The aim of 
the work was realization of excitation in a small volume, development of 
diagnostic methods for cold Rydberg atoms, spectroscopy of the microwave 
transitions between Rydberg states in the presence of the MOT quadrupole 
magnetic field, as well as observation of dipole-dipole interaction of a 
small number of Rydberg atoms.

\begin{figure*}
\includegraphics[scale=0.8]{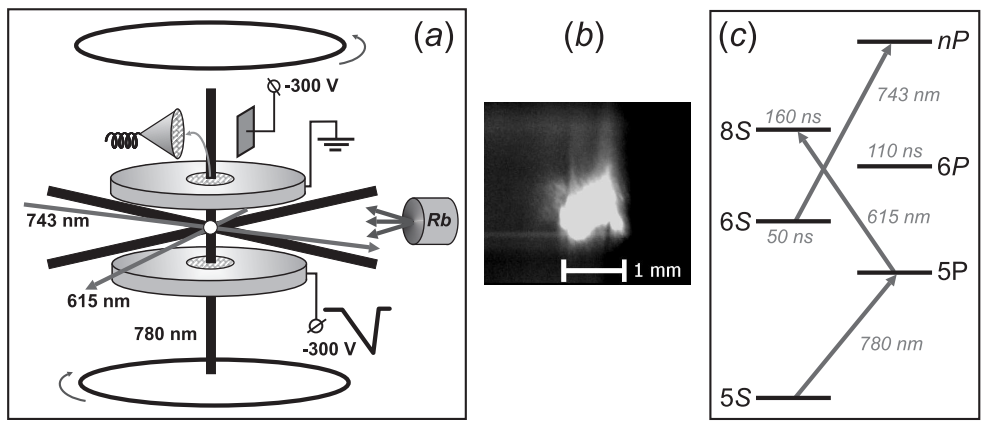}
\caption{\label{Fig1}(a) Schematic diagram of the experiment with cold Rb Rydberg atoms in a MOT. (b) Image of the cold Rb atom cloud after setting up the meshes on the vertical cooling laser beams (the shadows of the meshes with a cell size of 0.7 mm can be seen). (c) Energy level scheme for the three-step laser excitation of the Rb \textit{nP} Rydberg states.}
\end{figure*}

\section{Experimental setup}

The experiments were performed with cold Rb atoms in a MOT, illustrated in 
Fig.1(a). The MOT has a standard configuration [18] and consists of a 
vacuum chamber with optical windows, an oven as a Rb source, and 
anti-Helmholtz coils providing 3-dimensional magnetic field gradient of 
10$-$15~G/cm in the MOT center. The laser cooling system is based on two 
external-cavity diode lasers at the 780~nm wavelength. The atom cooling was realized by 
means of three pairs of mutually orthogonal light beams, each pair 
consisting of counter-propagating light waves of opposite circular 
polarization. One of the lasers (the cooling laser, with power 50~mW and 
linewidth of 1~MHz) was tuned to the closed $5S_{1/2} \left( {F = 3} \right) 
\to 5P_{3/2} \left( {F = 4} \right)$ transition of the $^{85}$Rb isotope, 
with a red detuning of 20$-$30~MHz. The second laser (repumping laser, with a 
power of 10~mW and linewidth of 1~MHz) was in resonance with the $5S_{1/2} 
\left( {F = 2} \right) \to 5P_{3/2} \left( {F = 3} \right)$ transition.

After adjustment and stabilization of the laser wavelength, a cloud of cold 
atoms of about 1~mm diameter was formed in the center of the MOT. The loading 
time of the MOT depended on the alignment of the laser beams and the Rb 
source temperature and was on the order of 1$-$3~s. The temperature of the 
trapped atoms was not measured, but it is known from literature that the 
typical values are 100$-$300~$\mu $K [18]. For example, in ref. [19] a similar 
MOT was reported and its temperature, measured using recoil-induced 
resonances, was about 140~$\mu $K.

The excitation and registration of Rydberg atoms was performed between the two 
stainless-steel plates, with a 10~mm hole in the center of each plate 
[Fig.1(a)]. In order to create a homogeneous electric field, these 
holes were closed with optically transparent (85\% transmission) metallic 
meshes. The distance between the plates was 10~mm. Electric field was used 
for detection of the Rydberg atoms by means of selective field ionization 
(SFI) [20]. The electrons formed after ionization were accelerated by the 
electric field, flied through the upper mesh, after which a deflecting 
electrode directed them towards the channeltron's input window. Its output 
pulses were processed by a fast analog-to-digital converter, gated 
integrator and a computer. This allowed us to control the number of atoms 
and the population of Rydberg states for a large range of the principal 
quantum number \textit{n}.

The transmission of the vertical cooling laser beams through the meshes 
leads to inhomogeneities of the intensity distribution and distortion of the 
cloud shape. The cloud image, taken with a CCD-camera, is shown in 
Fig.1(b). The shadows of the meshes with cell size about 0.7~mm are 
clearly seen. Nevertheless, in the center of the cloud the atomic 
distribution is homogeneous enough, which made possible to vary the position 
of excitation of the Rydberg atoms. In order to determine the total number 
of atoms in the cloud, the CCD camera was calibrated with respect to the 
intensity of the resonant fluorescence. Typically 10$^{5}-10^{6}$ atoms 
were observed, which corresponds to a density of the cold atoms of 
10$^{8}-10^{9}$ cm$ ^{-3}$.

The excitation of cold Rb atoms into the Rydberg states \textit{nP} 
(\textit{n}=30$ - $50) was performed in three steps [Fig.1(c)]. For 
the first step $5S_{1/2} \to 5P_{3/2} $ the cw cooling laser was used. The 
second step $5P_{3/2} \to 8S_{1/2} $ was excited using the radiation of a 
pulsed Rhodamine 6G dye-laser with a wavelength of 615~nm, pumped with the 
second harmonic of Nd:YAG laser at a pulse repetition rate of 5~kHz. The $8S$ 
state has a lifetime of 160~ns and decays into the lower P-states, including 
the $6P$-level. The latter has a lifetime of 110~ns and quickly decays into the 
$6S$ state having a lifetime of 50~ns. Calculations made in the kinetic-equations 
approximation have shown that about 10\% of the atoms from the $8S$ state 
cascade down to the $6S$ state. Further, as a third step, the Rydberg 
\textit{nP} states were excited from $6S$ by means of a pulsed Ti:Sapphire 
laser with a wavelength of 743~nm, pumped by the second harmonic of Nd:YAG 
laser at a pulse repetition rate of 5~kHz. Both pulsed lasers were 
synchronized in time and their pulse width was about 50~ns.

The radiations for the second- and third-step excitation were focused onto 
the cold atom cloud in crossed-beams geometry (Fig.1\textit{à}) by means 
of objectives with a focal length of 80~mm. The beam waist diameters measured 
at a level of 1/e$^{2}$ were $25 \pm 5$~$\mu $m for the dye-laser and $40 \pm 
5$~$\mu $m for the Ti:Sapphire laser. Thus, the intersection of the focused 
beams formed the effective excitation volume of Rydberg atoms whose 
dimensions were around 40$ - $100~$\mu $m, depending on the mutual position 
of the beam waists and on the presence or absence of transition saturation.

Microwave spectroscopy [20] was used for diagnostics of the cold Rb Rydberg 
atoms in the MOT. The microwave transitions between Rydberg states with 
\textit{n}=30-50 have natural linewidths on the order of several kHz and 
their frequencies are in the range of 30$-$100~GHz. As a source of microwave 
radiation we used a carcinotron oscillator G4-142, whose frequency with a 
linewidth of 20~kHz was locked to a quartz synthesizer and was scanned in the 
interval 53$ - $82~GHz. The radiation was introduced directly through the MOT 
window. Since the saturation intensity of the single-photon transitions is 
around 10$^{-12}-10^{-9}$~W/cm$^{2}$, they are easily excited 
with low-intensity radiation and detected with SFI. The microwave transition 
spectra provide information on the presence of external magnetic and 
electric fields and their spatial distribution. Such fields can be used for 
manipulating the energy levels of the Rydberg atoms.

The time diagram of the signals in the detection system is shown in Fig.2(a). Following each laser pulse exciting some of the cold atoms to 
an initial Rydberg state \textit{nP}, the atoms interacted with each other 
or with the microwave radiation for a time period \textit{t}$_{0}$=1$ - 
$10~$\mu $s. After that, a ramp of the ionizing electric field was applied 
with a rise time of about 2~$\mu $s. Depending on the Rydberg state of the 
atom, ionization took place at different moments in time after the laser 
pulse. The pulsed ionization signal was then registered at the output of 
VEU-6 using two gates, corresponding to the initial \textit{nP} and final 
\textit{n}$^\prime$\textit{L} state of the Rydberg atom. The number of electrons 
registered for a single laser pulse was determined by the number of Rydberg 
atoms in the excitation volume and by the detection efficiency of VEU-6 
[17]. 

Figure 2(b) shows a histogram of the amplitude of the output pulses 
from VEU-6. Several peaks can be seen, corresponding to the different 
numbers of registered Rydberg atoms (from 1 to 5). The mean number of atoms, 
detected per laser pulse, was 2.2. After each laser pulse, the data 
acquisition system measured the amplitude of the output VEU-6 signal at both 
registration channels (for the initial and final state), and determined the 
number of atoms according to the histogram, measured in advance. After 
accumulation of data for 1000$-$5000 laser pulses, it sorted the signals over 
the number of atoms and calculated the probability of the transition from 
the initial to the final Rydberg state.

\begin{figure}
\includegraphics[scale=0.7]{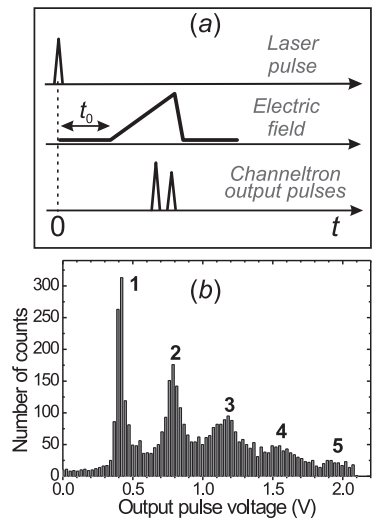}
\caption{\label{Fig2}(a) Time diagram of the pulses in the experiments with selective field ionization (SFI) of Rydberg atoms. (b) Histogram of the amplified output pulses of the channel multiplier VEU-6. }
\end{figure}

\section{Lifetime measurement}

First, a test measurement was performed to ensure that the registered 
signals are from the cold Rydberg atoms and not from the thermal atomic beam 
from the Rb source. For this, the effective lifetime $\tau _{\mathrm{eff}} $ of the 
Rydberg atoms in the MOT was determined. We have chosen as a test level the 
initial Rydberg state 37\textit{P}. The lifetime of the \textit{nP} states 
with other \textit{n} values can be estimated with the approximate scaling 
dependence $\tau _{\mathrm{eff}}\sim n^{3}$.

It should be noted that in atomic-beam experiments the measurement of the 
lifetime of high Rydberg states is not a trivial task, since the time of 
flight of the atoms through the detection system is usually much smaller 
than their lifetime. The situation with the cold Rydberg atoms in a MOT is quite opposite. For a temperature of the Rb cloud of 200~$\mu $K, the 
root-mean-square (rms) velocity of the atoms is 25~cm/s, so at a timescale of 
milliseconds the cloud does not expand significantly. Thus, direct 
measurement of the lifetime is possible even for states with \textit{n}$\sim 
$100. The first lifetime measurements of cold Rb Rydberg atoms are reported 
in ref. [21]. In a later work [22] it was proposed that the experiments in 
ref. [21] would yield overestimated values of the lifetimes.

In our experiments we measured the dependence of the mean number of Rydberg 
atoms in the 37\textit{P} state on the time delay of the SFI pulse (time 
\textit{t}$_{0}$, see Fig.2\textit{à}). Fig.3 shows the experimental 
records, made for cold atoms in the working MOT (black curve) and for hot 
atoms when the MOT magnetic field is off (grey curve). Both dependences are 
well described with the exponent $\mathrm{exp}\left( { - t_{0} /\tau _{\mathrm{eff}}}  
\right)$. The measured lifetime of the cold atoms was $\tau _{\mathrm{eff}} = \left( 
{38 \pm 3} \right)$~$\mu $s. This value is close to the previously calculated 
by us in ref. [23] value of 43~$\mu $s taking into account the lifetime 
decrease due to background blackbody radiation at 300~K. However, it 
noticeably differs from the experimental value of 47~$\mu $s reported in ref. 
[21], which could support the conclusions drawn in [22].

At the same time, for the hot atoms the effective lifetime was only $\tau 
_{\mathrm{eff}} = \left( {6 \pm 0.5} \right)$~$\mu $s. It is determined by the time of 
flight of the atoms through the detection system. Moreover, from Fig.3 it 
can be seen that the signal from the hot atoms is about 20 times weaker than 
the signal from the cold ones. Therefore, we could conclude that in a 
working MOT the contribution of hot atoms to the measured signal is less 
than 5\% and can be neglected.

\begin{figure}
\includegraphics[scale=0.5]{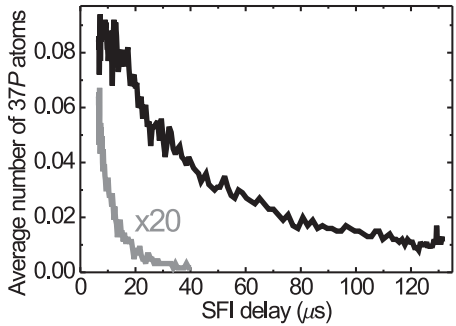}
\caption{\label{Fig3}Dependence of the mean number of atoms in the 37\textit{P} state on the time delay after the exciting laser pulse: black curve $-$ for cold Rydberg atoms in the working MOT, grey curve $-$ for hot atoms after switching off the MOT magnetic field.}
\end{figure}

\section{Influence of external fields}

Atoms in MOTs are trapped using an inhomogeneous quadrupole 
magnetic field created by anti-Helmholtz coils. Typical values of the field 
gradient are 10$ - $15~G/cm, thus the variation of the magnetic field within 
the atomic cloud of 1~mm in size is around 1~G. This field leads to Rydberg levels 
splitting of the order of 1$ - $3~MHz, which limits the spectral resolution 
in experiments with cold Rydberg atoms in a MOT. To overcome this, often the magnetic field is turned 
off prior to measurements [11,24]. However, the fall time for the field down 
to 50$ - $100~mG depends on the inductance of the coils and on the eddy 
currents in the metallic elements of the detection system and the vacuum 
chamber. Therefore it is difficult to be made less than several 
milliseconds, and typical values are 10$ - $20~ms. This leads to a  
decrease in the signal accumulation rate and to non-steadiness of the cold 
atomic cloud due to its gradual expansion. In addition, the slow switching 
of the magnetic field does not comply with the requirements to the quantum 
computer qubits, which should be manipulated on timescales of less than 
microsecond [16].

Therefore, in our experiments the magnetic field was not switched off. This 
provided high signals accumulation rate, determined by the exciting laser 
pulse rate (5~kHz). In order to improve the spectral resolution and to 
decrease the influence of the inhomogeneous magnetic field, we localized a 
small excitation volume of Rydberg atoms close to the center of the cold 
atomic cloud where the field turns to zero.

\begin{figure}
\includegraphics[scale=0.5]{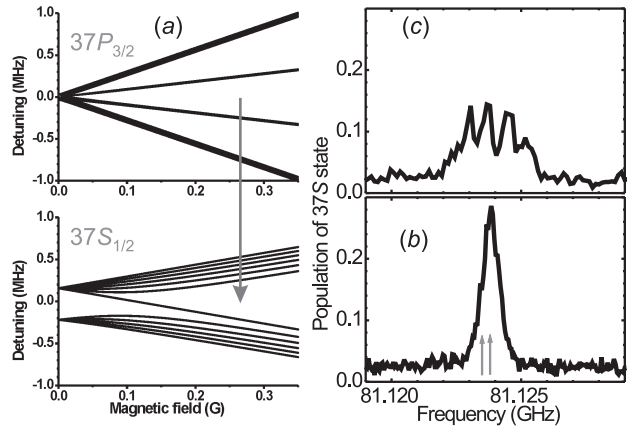}
\caption{\label{Fig4}(a) Energy level scheme for the microwave transition  $37P_{3/2} \to 37S_{1/2} $  in $^{85}$Rb atoms in a weak magnetic field, numerically calculated for the nonlinear Zeeman effect on the states  $37P_{3/2}$ and $37S_{1/2} $. (b) Spectrum of the microwave transition $37P_{3/2} \to 37S_{1/2} $ when the excitation volume is localized in the center of the cold atomic cloud and (c) at its periphery. The arrows indicate the calculated positions of the unresolved hyperfine-structure components.}
\end{figure}

The influence of the magnetic field was studied by microwave spectroscopy of 
the transition $37P_{3/2} \to 37S_{1/2} $ in $^{85}$Rb. The 
37\textit{S}$_{1/2}$ state has two hyperfine sublevels \textit{F}=2, 3 with 
hyperfine splitting of 376~kHz, and the 37\textit{P}$_{3/2}$ state has four 
hyperfine sublevels \textit{F}=1, 2, 3, 4 with a total width of 62~kHz (these 
values were calculated from the data for the lower-lying states using the 
scaling dependence $n^{-3}$). However, even a weak magnetic 
field can cause full mixing and splitting of the hyperfine levels, as can be 
seen from the calculated by us diagram in Fig.4(a). If the hyperfine 
structure is not taken into account, the energy shifts of the magnetic 
sublevels of the fine levels \textit{M}$_{J}$ are described with the simple 
formulas [25]:

\begin{equation}
\label{eq1}
\Delta E = g\mu _{B} BM_{J} ,\quad g\left( {S_{1/2}}  \right) = 2,\quad 
g\left( {P_{3/2}}  \right) = 4/3,
\end{equation}

\noindent
where \textit{g} is the Lande factor, $\mu _{B}$ $-$ the Bohr magneton, and 
\textit{B} $-$ the magnetic field. The values of \textit{g} were calculated 
using the general formula for the fine structure components of the 
\textit{S} and \textit{P} states [25].

The spectrum of the microwave transition $37P_{3/2} \to 37S_{1/2} $ was 
expected to be different for different positions of the excitation volume 
within the atomic cloud. This is confirmed by Fig.4, which shows the spectra 
obtained for single-atom signals when the excitation volume of about 100~$\mu 
$m was close to the center of the cloud (b) and to its periphery 
(c). In the center, the profile is almost Lorentzian, with a full 
width at half maximum (FWHM) of $670 \pm 30$~kHz. This width is determined 
mainly by the unresolved hyperfine structure of the 37\textit{S}$_{1/2}$ 
state and the time of interaction of the atoms with the microwave field 
(3~$\mu $s), which sets the Fourier-transform resonance width to about 
300~kHz. At the same time, at the cloud periphery the profile consists of 
four components with a total spectrum width of 2.7~MHz. The magnetic field 
magnitude, estimated using Eq.(\ref{eq1}), was $0.35 \pm 0.05$~G. Thus, such an 
experiment on localization of the Rydberg excitation into a small volume 
demonstrates a sort of "tomography" of the cold atomic cloud. It also 
confirms the possibility to obtain narrow resonances in an inhomogeneous 
magnetic field, under the condition that the excitation volume is localized 
around the zero magnetic field position. The latter is important for the 
provision of strong dipole-dipole interaction of Rydberg atoms, since even a 
weak magnetic field leads to decrease in the interaction energy by several times [11].

Rydberg atoms are also very sensitive to external electric fields, since the 
polarizability of the Rydberg states increases as \textit{n}$^{7}$. As an 
illustration, Fig.5 shows the spectrum of the microwave transition 
$37P_{3/2} \to 37S_{1/2} $ in a weak electric field. The \textit{S} and 
\textit{P} states of the Rb atoms are subject to quadratic Stark effect [see 
the scheme in Fig.6(a)]. The 37\textit{P}$_{3/2}$ state is split into 
two components with momentum projections $|M_{J}|$=1/2 and 
$|M_{J}|$=3/2. The splitting is determined by the tensor 
polarizability of the 37\textit{P}$_{3/2}$ level, which has been measured to 
be $\alpha _{2} \left( {37P_{3/2}}  \right) = - \left( {3.1 \pm 0.15} 
\right)$~MHz/(V/cm)$^{2}$. This value agrees well with our numerical 
calculations, which give $\alpha _{2} \left( {37P_{3/2}}  \right) = - 
3.2$~MHz/(V/cm)$^{2}$. At the same time, the level shifts are determined by 
the difference in scalar polarizability of the 37\textit{P}$_{3/2}$ and 
37\textit{S}$_{1/2}$ levels. The experimentally measured value of $\alpha 
_{0} \left( {37P_{3/2}}  \right) - \alpha _{0} \left( {37S_{1/2}}  \right) = 
- \left( {30 \pm 1.5} \right)$~MHz/(V/cm)$^{2}$ is also close to the 
calculated value of $-$29.5~MHz/(V/cm)$^{2}$.

It should be noted that the increase in the electric-field strength to 4.66~V/cm 
leads to resonance broadening by about 3~MHz. This is due to the 
electric field inhomogeneity over the excitation volume, which is relatively 
small $-$ only $\left( {0.5 \pm 0.1} \right)\% $, but still limits the 
spectral resolution.

\begin{figure}
\includegraphics[scale=0.7]{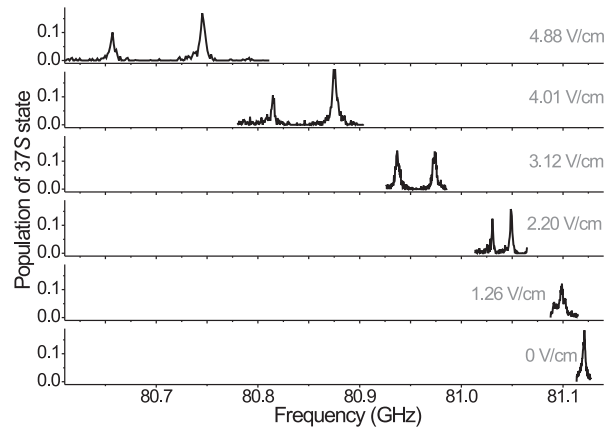}
\caption{\label{Fig5}Spectrum of the microwave transition $37P_{3/2} \to 37S_{1/2} $  in a weak electric field.}
\end{figure}

\section{Resonant dipole-dipole interaction}

As can be seen from the results of the previous experiment, it is possible 
to control the Rydberg state energies using a weak electric field. In 
particular, it can be used to Stark-tune the resonant dipole-dipole 
interaction of the Rydberg atoms [26], which is required for practical 
realization of quantum-logic operations and dipole blockade [6,27]. In the 
case of Rb atoms in \textit{nP} Rydberg state, the resonant interaction of 
two atoms is realized as [13,28]:

\begin{equation}
\label{eq2}  {{\begin{array}{*{20}l}
 {\mathrm{Rb}\left( {nP_{3/2}}  \right) + \mathrm{Rb}\left( {nP_{3/2}}  \right) \to \quad \quad \quad \quad }  \\ \\
 {\quad \quad \quad \quad \mathrm{Rb}\left( 
{nS_{1/2}}  \right) + \mathrm{Rb}\left( {\left( {n + 1} \right)S_{1/2}}  \right)\quad \quad } \\
\end{array}} 
}\end{equation}

The exact energy resonance for this process can be achieved by means of 
Stark tuning of the Rydberg levels in an electric field [Fig.6(a)]. 
It should be noted, though, that because of the specific values of the 
quantum defects and polarizabilities, resonance Stark tuning is possible 
only for states with $n \le 38$. Table 1 provides a list of the 
calculated by us initial detunings and mean critical values of the electric 
field for the states \textit{n}=35$-$40. In our experiment, the chosen initial 
state was 37\textit{P}$_{3/2}$, since for the higher state 
38\textit{P}$_{3/2}$ the critical electric field value for the double resonance 
is rather small, and thus the resonance can be broadened due to stray 
fields, while for lower states the frequencies of the microwave resonances 
are out of the working range of the microwave oscillator used for Rydberg-state diagnostics.

\begin{longtable}
{|p{70pt}|p{70pt}|p{70pt}|} a & a & a \kill \hline
\multicolumn{3}{|p{210pt}|}{\textbf{Table 1}. Calculated values of the initial detuning  $\Delta_{0} $ and critical field $E_{cr}$ for the
double Stark resonances from $nP$ states shown in Fig.6(a)
}  \\ \hline
\textit{n}\textit{}& $\Delta_{0} $ (MHz)\textit{}&
$E_{cr}$ (V/cm)\textit{} \\ \hline 35& 382& 4.5 \\ \hline
36& 228& 3.1 \\ \hline 37& 105& 1.9 \\ \hline
38& 5.6& 0.4 \\ \hline 39& -73&  \\ \hline 40& -136&  \\ \hline
\end{longtable}

At resonance the cross-section of the process (\ref{eq2}) strongly increases. As a 
result of the interaction, one of the atoms goes into the lower state 
\textit{nS}$_{1/2}$, and the other $-$ into the higher state 
(\textit{n}+1)\textit{S}$_{1/2}$. The atomic-state change is registered 
using the SFI method. As a rule, in the experiments the total population of 
the \textit{nS}$_{1/2}$ state is measured, since in the Rb atoms the 
(\textit{n}+1)\textit{S}$_{1/2}$ state has almost the same critical field 
for SFI as the \textit{nP}$_{3/2}$ state.  There can be more than two interacting atoms within the excitation volume, if they are closely spaced. In this 
case collective interactions in a cold Rydberg gas take place [3].

As mentioned earlier, the specific feature of our experiments is the 
possibility to determine the number of Rydberg atoms \textit{N} and their 
states after each laser pulse. The signals from 1000$-$5000 pulses are 
accumulated and subsequently automatically sorted over the number of atoms, 
and the transition probability for each group is calculated. As a result, 
the following signals are measured:

\begin{equation}
\label{eq3}
S_{N} = \frac{{n_{N} \left( {37S} \right)}}{{n_{N} \left( {37P} \right) + 
n_{N} \left( {37S} \right) + n_{N} \left( {38S} \right)}} \quad.
\end{equation}

\noindent Here $n_{N} \left( {nL} \right)$ is the total number of detected Rydberg 
atoms in a state \textit{nL} for the case of \textit{N} Rydberg atoms. 
In fact, the signal \textit{S}$_{N}$ describes the mean transition 
probability for each atom after interaction with the $N-1$ 
surrounding atoms.

The resonant dipole-dipole interaction is registered through the dependence 
of the signals \textit{S}$_{N}$ on the strength of the dc electric field. In 
zero field, the signals have a small background related to the non-resonant 
transitions induced by blackbody radiation and atomic collisions. When the 
electric field is tuned to resonance, the signal increases due to the 
increased cross-section of the process (\ref{eq2}). The amplitude and width of the 
resonances for cold Rydberg atoms depend on the energy and the duration of 
the dipole-dipole interaction.

The operator of the dipole-dipole interaction between two atoms \textit{a} 
and \textit{b} is written as:

\begin{equation}
\label{eq4}
\hat {V}_{ab} = \frac{{1}}{{4\pi \varepsilon _{0}} }\left[ {\frac{{\hat 
{\mathbf{d}}_{a} \hat {\mathbf{d}}_{b}} }{{R_{ab}^{3}} } - \frac{{3\,\,\left( {\hat {\mathbf{d}}_{a} 
\mathbf{R}_{ab}}  \right)\,\left( {\hat {\mathbf{d}}_{b} \mathbf{R}_{ab}}  \right)}}{{R_{ab}^{5}} }} 
\right].
\end{equation}

\noindent Here $\hat {\mathbf{d}}_{a} $ and $\hat {\mathbf{d}}_{b} $ are the dipole moment operators for 
the atoms \textit{a} and \textit{b}, $\mathbf{R}_{ab} $ is a vector connecting the 
two atoms, and $\varepsilon _{0}$ $-$ the dielectric constant. To calculate 
the evolution of the Rydberg state population, it is necessary to solve the 
problem for a quasimolecule formed by the two atoms [13,17]. For motionless 
Rydberg atoms this problem can be solved analytically. The resulting 
evolution of the population of the final state 37\textit{S} in each atom is 
described by the following expression:

\begin{equation}
\label{eq5}
\rho _{2} \left( {t_{0}}  \right) \approx \frac{{1}}{{2}}\frac{{2\Omega 
_{ab}^{2}} }{{2\Omega _{ab}^{2} + \Delta ^{2}/4}}\mathrm{sin}^{2}\left( {\sqrt 
{2\Omega _{ab}^{2} + \Delta ^{2}/4} \;t_{0}}  \right),
\end{equation}

\noindent
where $\Omega _{ab} = V_{ab} /\hbar $ is the dipole-dipole interaction 
energy in frequency units, and $\Delta = \left( {2\mathrm{E}_{37P} - \mathrm{E}_{37S} - 
\mathrm{E}_{38S}}  \right)/\hbar $ is the detuning from exact resonance. Thus, the 
interaction between two motionless Rydberg atoms leads to oscillations in 
the final state population of each atom, in a way similar to the Rabi 
oscillations in a two-level atom. Since this process is coherent, it can be 
used for realization of two-qubit logical operations and of quantum logic 
gates based on neutral atoms.

\begin{figure*}
\includegraphics[scale=0.8]{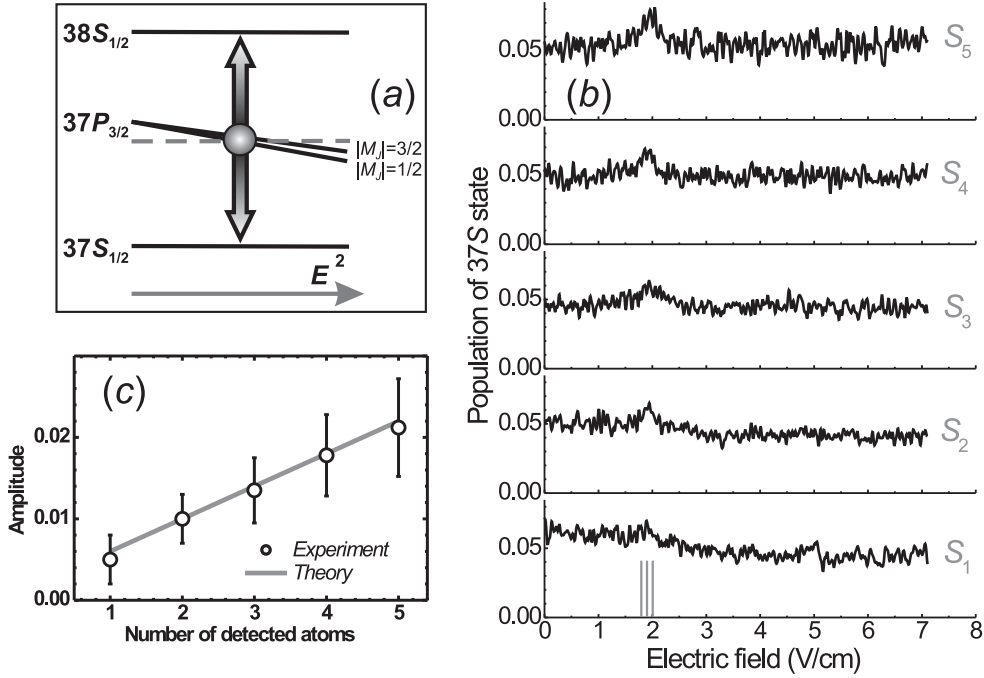}
\caption{\label{Fig6}(a) Scheme of the appearance of the double Stark resonance   $37S_{1/2} - 37P_{3/2} - 38S_{1/2} $ in the electric field $E=1.8-2.0$~V/cm. (b) Spectra of the resonant dipole-dipole interaction for selectively detected 1 to 5 Rydberg atoms. The vertical lines denote the calculated positions of the resonances. (c) Dependence of the resonance amplitude on the number of detected Rydberg atoms. The dots represent experimental data, and the line is theoretical dependence for weak dipole-dipole interaction.}
\end{figure*}

It is impossible to obtain a general analytical formula for the transition 
probability $\rho _{N} \left( {t_{0}}  \right)$ for a larger number of 
interacting atoms. This is due to the fact that besides the resonant 
dipole-dipole interaction (\ref{eq2}), it is necessary to take into account the 
exchange processes like $\mathrm{Rb}\left( {37P} \right) + \mathrm{Rb}\left( {37S} 
\right) \to \mathrm{Rb}\left( {37S} \right) + \mathrm{Rb}\left( {37P} \right)$ and $\mathrm{Rb}\left( 
{37P} \right) + \mathrm{Rb}\left( {38S} \right) \to \mathrm{Rb}\left( {38S} \right) + \mathrm{Rb}\left( 
{37P} \right)$, which are always resonant and lead to a population exchange 
between neighboring atoms [3]. If these processes are not taken into 
account, then Eq.(\ref{eq5}) should be simply modified by substituting $2\Omega 
_{ab}^{2} $ with the sum of the squares of the interaction energies for all 
atomic pairs, and by replacing the scaling factor 1/2 by 1/\textit{N}: 

\begin{equation}
\label{eq6}  {{\begin{array}{*{20}l}
 {\rho _{N} \left( {t_{0}}  \right) \approx 
\displaystyle\frac{{1}}{{N}}\frac{{\sum\limits_{a \ne b} {\Omega _{ab}^{2}}  
}}{{\sum\limits_{a \ne b} {\Omega _{ab}^{2}}  + \Delta ^{2}/4}} \quad \times \quad \quad \quad \quad }  \\ \\
 {\quad \quad \quad \quad \mathrm{sin}^{2}\left( 
{\sqrt {\sum\limits_{a \ne b} {\Omega _{ab}^{2}}  + \Delta ^{2}/4} \;t_{0}}  
\right)\quad \quad } \\
\end{array}} 
}\end{equation}

\noindent Here the total transition probability is calculated as the sum of the 
probabilities over all possible binary interactions. Such an approximation 
is valid for a weak dipole-dipole interaction, when the \textit{S}-states 
are weakly populated and $\rho _{N} \left( {t_{0}}  \right) < < 1$. The 
numerical modeling which takes into account the exchange interactions 
confirms the applicability of Eq.(\ref{eq6}) for this case.

The signals \textit{S}$_{N}$ measured in our experiment, however, cannot be 
directly modeled with Eq.(\ref{eq6}). First of all, this expression is obtained for 
the two motionless atoms with a fixed separation and fixed dipole orientation. 
In our experiment the signals are accumulated from Rydberg atoms with 
arbitrary position within the excitation volume and with arbitrary dipole 
orientations. Therefore, it is necessary to average Eq.(\ref{eq6}) over interatomic 
distances and dipole orientations. Second, as was shown in our earlier work 
[17], due to the finite detection probability of the SFI detector, the 
signals \textit{S}$_{N}$, corresponding to detection of \textit{N} 
Rydberg atoms, are a mixture of signals from a larger number of actually 
excited atoms $i \ge N$:

\begin{equation}
\label{eq7}
S_{N} = \,\mathrm{e}^{ - \bar {n}\left( {1 - T} \right)}\sum\limits_{i = N}^{\infty}  
{\rho _{i} \left( {t_{0}}  \right)\frac{{\left[ {\bar {n}\left( {1 - T} 
\right)} \right]^{i - N}}}{{\left( {i - N} \right)!}}} ,
\end{equation}

\noindent
where $\bar {n}$ is the mean number of Rydberg atoms excited per laser 
pulse, and \textit{T} $-$ the detection efficiency of the SFI detector.

Nevertheless, it is possible to use Eq.(\ref{eq6}) for estimation of the resonance 
amplitude and width, if the mean distance between the Rydberg atoms in the 
excitation volume is introduced and the dipole orientation is not considered 
(scalar model). Since the summation in Eq.(\ref{eq6}) is over $N\left( {N - 1} \right)$ 
terms, we can write:

\begin{equation}
\label{eq8}
\sum\limits_{a \ne b} {\Omega _{ab}^{2} \equiv N\left( {N - 1} \right)\,\bar 
{\Omega} ^{2}} ,
\end{equation}

\noindent
where $\bar {\Omega} $ is the rms energy of the dipole-dipole interaction of 
the two Rydberg atoms with arbitrary positions within the excitation volume. For 
weak interaction Eq.(\ref{eq6}) yields for the resonance amplitude:

\begin{equation}
\label{eq9}
\rho _{N} \left( {t_{0} ,\Delta = 0} \right) \approx \left( {N - 1} 
\right)\bar {\Omega} ^{2}t_{0}^{2} ,
\end{equation}

\noindent In this case the resonance width is the same for all \textit{N} and is 
determined only by the interaction time \textit{t}$_{0}$. Substituting (\ref{eq9}) 
into (\ref{eq7}) and summing over \textit{i}, we obtain:

\begin{equation}
\label{eq10}
S_{N} \approx \left[ {N - 1 + \;\bar {n}\left( {1 - T} \right)} \right]\bar 
{\Omega} ^{2}t_{0}^{2} ,
\end{equation}

Thus, at weak dipole-dipole interaction the \textit{S}$_{N}$ dependence on 
\textit{N} should be linear. At \textit{N}=1 the signal \textit{S}$_{1}$ is 
due to the imperfection of the SFI detector, and it characterizes the 
measurement error in the experimental signals.

In our experiment the observation of dipole-dipole interaction was 
performed at the maximal concentration of the cold atoms of 10$^{9}$~cm$ 
^{-3}$. The excitation volume of about 100~$\mu $m in size was localized 
close to the center of the cold atomic cloud. The time of free interaction 
of the Rydberg atoms was $t_{0} \approx 2.2$~$\mu $s. The resonant 
dipole-dipole interaction spectra were obtained for 1 to 5 registered 
Rydberg atoms. These spectra are presented in Fig.6(b). The vertical 
lines indicate the calculated positions of the three possible resonances 
occurring at 1.79, 1.89 and 2.0~V/cm for the different Stark components of 
the 37\textit{P}$_{3/2}$ state. As can be seen from the experimental 
spectra, the individual resonances overlap and form a single resonance with 
a width of about 250~mV/cm. Presumably, the resonance broadening was due to 
the electric field inhomogeneity over the excitation volume (10$-$20~mV/cm), as 
well as due to the stray ac fields whose amplitude was estimated to be up to 
30$-$50~mV/cm.

The broadening of the resonance lead to the low signal amplitude and 
signal-to-noise ratio, as seen from Fig.6(b). Nevertheless, it was 
possible to measure the amplitude versus the number of detected Rydberg 
atoms [Fig.6(c)]. The obtained dependence is linear, which agrees 
well with the theoretical prediction of Eq.(\ref{eq10}).

Moreover, using the experimental dependence shown in Fig.6(c) and the 
method proposed by us in [17], it was possible to determine the unknown 
detection efficiency of Rydberg atoms $T = \left( {40 \pm 15} \right)\% $ 
and the mean number of excited Rydberg atoms $\bar {n} = \left( {1.7 \pm 
0.5} \right)$. This measurement was based on the measured amplitude ratio of the 
two- and one-atom resonances $S_{2} /S_{1} \approx 2 \pm 0.7$. The mean 
number of registered atoms in the present experiment $\bar {n}T = 0.7 \pm 
0.1$, needed for the above estimations, was experimentally obtained 
from a histogram similar to that presented in Fig.2(b).

The above experimental values and Eq.(\ref{eq10}) make possible to estimate 
the unknown parameter $\bar {\Omega} ^{2}t_{0}^{2} = 0.004 \pm 0.0015$. 
Thus, for an interaction time of 2.2~$\mu $s we can estimate the rms energy 
of the dipole-dipole interaction of the two atoms in the excitation volume to be 
$\bar {\Omega} /2\pi = 4.7 \pm 1.6$~kHz. This value is close to the 
theoretical estimation of 3~kHz, obtained for a mean distance between the 
atoms $R = 50$~$\mu $m and for radial parts of the dipole moments 1320 and 
1290~a.u. of the transitions $37P_{3/2} \to 37S_{1/2} $ and $37P_{3/2} \to 
38S_{1/2} $, respectively, which were calculated numerically by us.

It should be noted, that the increase in supply voltage of VEU-6 up to 3.5~kV 
lead to increase in \textit{T} of three times as compared to work [17], 
where the supply voltage was 2.5~kV. The high detection probability is a necessary 
prerequisite in the experiments on quantum logic gates, since the 
measurement error depends on $\bar {n}\left( {1 - T} \right)$, as can be 
seen from Eq.(\ref{eq10}).

Further experiments on the investigation of the dipole-dipole interaction of 
a small number of Rydberg atoms in an excitation volume of 20$-$30 $\mu $m are 
of interest.  It is expected that the decrease in excitation volume and better shielding against stray electric fields would allow for the observation of narrower resonances due to the decrease in magnetic and electric field inhomogeneity, as well as would increase the resonance amplitude due to the decrease in the mean distance between the atoms.

\section{Conclusion}

In this paper we presented the results of our first experiments with cold Rb 
Rydberg atoms in a MOT. The experiments have revealed the main features of 
the excitation and detection of Rydberg atoms in a small excitation volume 
within the cold atom cloud. It has been shown that due to the cooling and 
trapping of atoms, the effective lifetime of the cold Rydberg atoms in a MOT 
is close to the natural one. This makes possible to perform measurements at 
a large timescale, taking all advantages of Rydberg atoms.

The localization in a small excitation volume allows to select the position 
of atomic interaction within the inhomogeneous external fields, and to map 
the spatial distribution of these fields using microwave spectroscopy. In 
particular, localizing the excitation volume close to the zero magnetic-field point improves the spectral resolution and yields narrow microwave 
resonances even without turning off the MOT quadrupole magnetic field.

The main characteristic feature of the present experiment was the detection 
of Rydberg atoms by means of a channel electron multiplier. This enables the 
sorting of the obtained signals over the number of detected atoms for each 
laser pulse. As a result, it was possible to measure the dependence of the 
amplitude of the dipole-dipole interaction resonance on the number of 
Rydberg atoms, which agrees well with the theory for weak dipole-dipole 
interaction. The obtained results are of interest for the development of 
quantum logic gates based on cold Rydberg atoms.

\begin{acknowledgments}
The authors are grateful to Christina Andreeva for fruitful discussions and 
help in preparing this paper. The work was supported by RFBR (grant 
No.~05-02-16181) and by the Russian Academy of Sciences.
\end{acknowledgments}

\end{document}